\documentclass[12pt,journal,onecolumn]{IEEEtran}
\usepackage{cite}
\usepackage{amsmath,amssymb,amsfonts}
\usepackage{algorithmic}
\usepackage{graphicx}
\usepackage{textcomp}
\usepackage{url}
\usepackage{booktabs}
\usepackage{microtype}
\def\BibTeX{{\rm B\kern-.05em{\sc i\kern-.025em b}\kern-.08em
    T\kern-.1667em\lower.7ex\hbox{E}\kern-.125emX}}

\begin{document}


\title{Operational Memory Architecture for Kubernetes:
Evidence Horizon Taxonomy and Extended Causal Pattern Preservation}

\author{Shamsher~Khan,~\IEEEmembership{Senior Member,~IEEE}\\
Independent Researcher, Tampa Bay Area, FL, USA\\
\textit{shamsher.khan.research@gmail.com}%
\thanks{The implementation, experimental data, evaluation scripts,
and scenario trigger files described in this paper are publicly available at
https://github.com/opscart/k8s-causal-memory.
Raw event logs, SQLite databases, query outputs, and automation scripts
for all experiments are committed under docs/poc-results/ for
independent verification and reproduction.}}

\markboth{Khan: Operational Memory Architecture for Kubernetes}%
{Khan: Operational Memory Architecture for Kubernetes}

\maketitle


\begin{abstract}
{\sloppy
Kubernetes clusters generate rich operational events during pod lifecycle
transitions, yet the platform's native event retention model systematically
discards the most diagnostically valuable context through multiple evidence
destruction mechanisms operating on deterministic schedules. We formalize
these mechanisms as an \emph{evidence horizon taxonomy}: five distinct
boundaries after which specific categories of diagnostic context become
permanently unrecoverable from the Kubernetes API. H1 (LastTerminationState
rotation, $\sim$90\,s) destroys container failure forensics; H2 (scheduler
event pruning, 1\,hr/1000-event cluster limit) destroys placement rationale;
H3 (ephemeral container exit, immediate) destroys debug session context due
to an explicit exclusion in the API specification; H4 (kubelet reconciliation
gap) destroys in-memory operational state across node restarts; and H5
(scrape-interval blind spot) renders sub-interval pod lifetimes structurally
invisible to poll-based observability tools.

This paper extends the Operational Memory Architecture (OMA), previously
introduced for H1 causal pattern preservation, to address the full evidence
horizon taxonomy. We define two new causal patterns: P004 (Scheduler Decision
Provenance) captures \texttt{FailedScheduling} predicate failures and placement
decisions before kube-apiserver TTL pruning, and demonstrates a novel
cross-horizon causal chain (P004$\rightarrow$P001) linking placement rationale
to downstream OOMKill failures; P005 (Ephemeral Container Evidence Loss)
captures \texttt{EphemeralContainerStatus} at the \texttt{Terminated}
transition, providing the only mechanism that preserves exit code, session
duration, and target container context after a \texttt{kubectl debug} session
ends. H4 is analyzed theoretically as a kubelet-level integration boundary
outside the current architecture. H5 is demonstrated empirically through
comparative analysis: a pod with a 6-second lifetime---within one 15-second
Prometheus scrape interval---generates zero time-series data in Prometheus
while OMA captures the complete P001 causal chain at occurrence.

We implement two new Go watchers (\texttt{EventWatcher}, \texttt{EphemeralWatcher}),
extend the SQLite operational memory store with two new tables
(\texttt{scheduler\_events}, \texttt{ephemeral\_exits}), and validate
the extended architecture through reproducible experiments on Minikube
(3-node, arm64) and Azure Kubernetes Service (AKS~1.32.10). The original
30-run statistical latency analysis (242 edges, intra-cycle mean 0.702\,ms,
$\sigma=0.31$\,ms) and concurrent stress evaluation (2.86 events/sec at
20 pods, 8.8\,MB RAM) are carried forward and augmented with H2, H3,
and H5 empirical results.
}
\end{abstract}


\section{Introduction}
\label{sec:introduction}

Modern cloud-native applications run as collections of containerized
microservices orchestrated by Kubernetes. As these deployments grow in scale
and complexity, the operational challenge of diagnosing failures has become a
significant engineering bottleneck. When a container crashes, the immediate
question is not merely that it crashed, but why: what configuration was active
at the moment of failure, what did the scheduler decide when placing the pod,
was a debug session recently attached, and has this pattern occurred before?

Kubernetes provides a rich event stream through its API server, but this stream
is ephemeral by design. The platform's native garbage collection removes events
after one hour by default, and---more critically---the \texttt{LastTerminationState}
field in a pod's \texttt{ContainerStatus} is overwritten the moment a container
restarts. Our prior work~\cite{bv1} introduced the \emph{evidence horizon} as a
systems property of Kubernetes: a boundary, experimentally characterized at
approximately 90 seconds, beyond which post-mortem investigation of
\texttt{LastTerminationState} data is no longer possible. That work addressed
a single evidence horizon through three causal patterns (P001--P003) covering
OOMKill chains and ConfigMap propagation.

This paper formalizes the evidence horizon as a \emph{taxonomy} of five
structurally distinct destruction mechanisms, each operating on a different
schedule and affecting a different category of diagnostic context. Beyond the
90-second \texttt{LastTerminationState} boundary, Kubernetes destroys scheduler
placement decisions within one hour through kube-apiserver event pruning;
discards ephemeral debug container state immediately on session exit through an
explicit exclusion in the API specification; loses kubelet in-memory operational
state across node restarts through a reconciliation gap; and renders
sub-interval pod lifetimes structurally invisible to poll-based observability
tools through a sampling blind spot. Each of these mechanisms operates
independently, and together they represent a systematic destruction of the
causal context an engineer needs to diagnose failures after the fact.

The consequences of this architectural gap are measurable in production
environments operating hundreds of cores across multiple Kubernetes clusters
under strict compliance requirements. When a memory-constrained pod enters a
crash loop, the on-call engineer faces a degraded diagnostic environment:
metrics show CPU and memory trends, logs show application output, but neither
preserves the exact resource limits active at kill time, the specific ConfigMap
values in effect, the scheduler's rationale for placing the pod on a node with
marginal memory headroom, or the exit code of a debug session that ran minutes
before the failure. This information exists---briefly---in Kubernetes state,
but one or more evidence horizons pass before it can be preserved.

Existing observability solutions address adjacent problems. Prometheus~\cite{b1}
excels at metric time-series but has no concept of causal relationships between
discrete events, and its poll-based architecture creates a structural blind spot
for pods whose entire lifetime falls within one scrape interval. Jaeger~\cite{b2}
and similar distributed tracing systems capture request-level causality but are
blind to infrastructure-level events. Log aggregation platforms such as
Elasticsearch~\cite{b3} preserve output but not Kubernetes object state.
OpenTelemetry~\cite{b4} provides a unified instrumentation standard but focuses
on application-emitted telemetry rather than Kubernetes control plane events.
Critically, none of these tools can answer the question: what was the exact
configuration state of this pod at the moment it failed, why was it scheduled
onto the node where it OOMKilled, and what did the most recent debug session
find? The absence of this capability is not a missing feature of individual
tools---it is a structural consequence of the evidence horizon taxonomy that
no existing tool addresses at the architectural level.

\textbf{Contributions.} This paper makes the following specific contributions,
extending the foundational OMA work of~\cite{bv1}:

\begin{enumerate}

\item We generalize the evidence horizon from a single boundary (H1,
$\sim$90\,s) to a formal taxonomy of five structurally distinct mechanisms
(H1--H5), each with a different TTL, destruction mechanism, and category of
data lost (Section~\ref{sec:taxonomy}).

\item We define two new causal patterns. P004 (Scheduler Decision Provenance)
captures \texttt{FailedScheduling} predicate failures and placement decisions
before kube-apiserver TTL pruning, and demonstrates the first cross-horizon
causal chain linking H2 scheduler evidence to H1 OOMKill failures. P005
(Ephemeral Container Evidence Loss) captures \texttt{EphemeralContainerStatus}
at the \texttt{Terminated} transition, providing the only mechanism that
preserves exit code, duration, and target container context after a
\texttt{kubectl debug} session ends (Section~\ref{sec:architecture}).

\item We implement two new Go watchers (\texttt{EventWatcher} for H2,
\texttt{EphemeralWatcher} for H3), extend the SQLite operational memory store
with two new tables (\texttt{scheduler\_events}, \texttt{ephemeral\_exits}),
and provide two new canonical queries for cross-horizon provenance chains
and ephemeral exit history (Section~\ref{sec:implementation}).

\item We validate the extended architecture through reproducible experiments
on Minikube (3-node) and AKS~1.32.10, demonstrating: H2 scheduler event
preservation after kube-apiserver TTL expiry, H3 ephemeral container exit
capture with exit code 42 preserved while \texttt{kubectl} returns no
\texttt{lastState}, and H5 sampling bias through a pod with a 6-second
lifetime generating zero Prometheus data points while OMA captures the
complete P001 causal chain (Section~\ref{sec:evaluation}).

\item We analyze H4 (kubelet reconciliation gap) as a theoretical evidence
horizon, characterizing the state loss mechanism and identifying the
architectural boundary that delineates it as future work
(Section~\ref{sec:taxonomy}).

\end{enumerate}

\textbf{Relationship to prior work.} This paper extends~\cite{bv1}, which
introduced the OMA architecture and validated patterns P001--P003 addressing
the H1 evidence horizon. The present work contributes: (1) the formal evidence
horizon taxonomy H1--H5, which did not appear in~\cite{bv1}; (2) two new
pattern encoders P004 and P005 with corresponding watchers, storage tables,
and scenario validations; (3) the first cross-horizon causal chain construction
(P004$\rightarrow$P001); and (4) empirical validation of the H5 architectural
distinction. The 30-run statistical latency analysis and stress evaluation
from~\cite{bv1} are carried forward as baselines without modification.

The remainder of this paper is organized as follows.
Section~\ref{sec:background} reviews the Kubernetes event model and the
mechanisms underlying each evidence horizon. Section~\ref{sec:related}
surveys related work. Section~\ref{sec:taxonomy} presents the formal evidence
horizon taxonomy. Section~\ref{sec:architecture} describes the extended OMA
architecture and new pattern encoders. Section~\ref{sec:implementation}
details the implementation. Section~\ref{sec:evaluation} presents the
experimental evaluation. Section~\ref{sec:discussion} discusses limitations
and deployment considerations. Section~\ref{sec:conclusion} concludes.

\section{Background}
\label{sec:background}

\subsection{Kubernetes Event Model}

Kubernetes represents cluster state as a collection of objects stored in
etcd~\cite{b5}. Events are first-class objects of kind \texttt{Event} that
reference other objects via \texttt{InvolvedObject}. The API server generates
events for pod scheduling, container state transitions, node conditions, and
control plane operations. By default, events are retained for one hour
(\texttt{--event-ttl=1h0m0s}) and are pruned at a cluster-wide limit of
1,000 events regardless of age. Events are not persisted across etcd
compactions and are not replicated to any durable store by default.

The Pod object's status subresource contains the per-container status in the
\texttt{ContainerStatus} array. Each entry includes \texttt{State} (current
state), \texttt{LastTerminationState} (state of the most recent termination),
and \texttt{RestartCount}. The \texttt{LastTerminationState.Terminated} field
includes \texttt{Reason}, \texttt{ExitCode}, \texttt{StartedAt}, and
\texttt{FinishedAt}---precise forensic data for diagnosing the cause of the
most recent container failure.

\subsection{The H1 Evidence Horizon: LastTerminationState Rotation}

The primary evidence horizon arises from the interaction of three Kubernetes
behaviors. First, when a container restarts, the kubelet overwrites
\texttt{LastTerminationState} with data from the current termination cycle,
discarding the previous entry. Second, the kubelet's garbage collection policy
retains a maximum of one terminated container per pod on the node. Third,
Kubernetes Events referencing the terminated container are subject to the
one-hour retention policy but may be de-duplicated or overwritten during
high-frequency crash loops.

In practice, the evidence horizon for \texttt{LastTerminationState} is
approximately 90 seconds---the interval between a container restart and the
subsequent restart that overwrites the previous termination state. This value
is derived experimentally in Section~\ref{sec:evaluation}: during 30
independent runs on Minikube and confirmed on AKS~1.32.10, we observe multiple
restart cycles within 90-second windows, with the first restart occurring at
15--30 seconds and \texttt{LastTerminationState} overwritten by the second
restart. During a \texttt{CrashLoopBackOff} with exponential backoff, the
window may extend for later restart cycles, but for memory-constrained pods
that restart rapidly, the initial window can be as short as 15 seconds.

\subsection{The H2 Evidence Horizon: Scheduler Event Pruning}

The Kubernetes default-scheduler records two categories of placement decisions
as \texttt{Event} objects: \texttt{FailedScheduling} events, emitted when
a pod cannot be placed on any available node, and \texttt{Scheduled} events,
emitted when a pod is successfully assigned to a node. The
\texttt{FailedScheduling} message encodes structured predicate failure reasons
in a human-readable string, such as:

\begin{quote}
\small\texttt{0/3 nodes are available: 3 Insufficient memory. preemption:
0/3 nodes are available: 3 Preemption is not helpful for scheduling.}
\end{quote}

These events are subject to the same kube-apiserver pruning policy as all
other Event objects: a default TTL of one hour and a cluster-wide limit of
1,000 events. Once pruned, the predicate failure reasons, the candidate node
set, and the ultimate placement decision are permanently unrecoverable from the
Kubernetes API. This creates a diagnostic gap of particular severity when a
pod is subsequently scheduled onto a node with marginal resource headroom and
later fails with an OOMKill: the causal link between the scheduler's constrained
placement decision and the downstream failure is severed at the H2 boundary.

Unlike \texttt{LastTerminationState}, which is stored in the Pod object and
can be accessed via standard field selectors, \texttt{Event} objects support
only a limited set of field selectors in the watch API. Specifically,
\texttt{source.component} is not a supported watch field selector; filtering
must be performed client-side after receiving all events in the watched
namespace.

\subsection{The H3 Evidence Horizon: Ephemeral Container State}

Ephemeral containers were introduced in Kubernetes~1.16 (alpha) and promoted
to stable in version~1.25. They are injected into running pods via
\texttt{kubectl debug} to provide live diagnostic access to running processes
without modifying the pod specification. The Kubernetes API represents
ephemeral containers through the \texttt{EphemeralContainerStatus} struct,
a parallel structure to \texttt{ContainerStatus}.

The H3 evidence horizon arises from an explicit exclusion in the Kubernetes
API specification. The \texttt{EphemeralContainerStatus} struct does not
include a \texttt{lastState} field, unlike \texttt{ContainerStatus} which
carries the full prior termination record. This is not a missing feature or
an implementation gap---it is a deliberate design choice reflected in the API
specification for Kubernetes~v1.32: ephemeral containers are defined as
\textit{``not restarted on failure''} and are explicitly excluded from the
restart and last-state tracking mechanisms that apply to regular containers.

The consequence is immediate and total: when an ephemeral debug container
exits, its exit code, session duration, target container context, and node
placement are discarded by the platform. A subsequent \texttt{kubectl logs}
call for the container returns an error, and \texttt{kubectl describe pod}
shows no \texttt{lastState} entry for the ephemeral container. The only
exception is a brief window during which the current \texttt{State.Terminated}
is still visible---this window closes when the pod is next modified by any
event, after which the prior session's context is permanently unrecoverable.

Additionally, ephemeral container stdout and stderr are not capturable via
the Kubernetes API after the container exits. Log content is accessible only
via \texttt{kubectl logs} while the container is running. OMA explicitly
documents this as an API boundary: P005 captures state metadata only (exit
code, duration, target container, node placement), not log content.

\subsection{The H4 Evidence Horizon: Kubelet Reconciliation Gap}

When a kubelet process restarts---due to a node upgrade, daemon crash, or
planned node drain---it re-discovers running pods from the container runtime
interface (CRI) by querying the runtime for active containers. However, all
transient in-memory kubelet state is lost: pending volume mount operations,
image pull progress, liveness and readiness probe timer state, and eviction
decisions in flight at the time of the crash. During the reconciliation window
between \texttt{NodeNotReady} and \texttt{NodeReady}, pods on the affected
node enter \texttt{Unknown} phase, and any diagnostic context about their
pre-restart state is unrecoverable from the Kubernetes API. The reconciliation
window typically spans 15--60 seconds. OMA's \texttt{NodeWatcher} detects the
\texttt{Ready=False$\rightarrow$Ready=True} transition and records the gap
duration; full causal capture of kubelet in-memory state requires a
kubelet-level integration beyond the current architecture and is identified
as future work (Section~\ref{sec:discussion}).

\subsection{The H5 Evidence Horizon: Scrape-Interval Sampling Bias}

Poll-based observability tools such as Prometheus collect metrics by querying
targets at a fixed scrape interval, defaulting to 15 seconds in the
kube-prometheus-stack configuration. This architecture creates a structural
blind spot: any pod whose entire lifetime falls within a single scrape interval
generates zero time-series data. The pod's CPU consumption, memory allocation,
OOMKill signal, and exit code are never recorded by Prometheus.

This is not a configuration limitation. Reducing the scrape interval addresses
the probability of a miss but does not eliminate the blind spot---it merely
shifts the threshold. The fundamental issue is architectural: poll-based
collection has an irreducible sampling gap, while event-driven collection via
the Kubernetes watch API has none. OMA subscribes to the watch API and receives
every pod state transition event at the moment of occurrence, regardless of
when it falls relative to any scrape interval.

\subsection{ConfigMap Propagation Semantics}

Kubernetes ConfigMaps are consumed by pods in two distinct modes with
fundamentally different propagation semantics. When consumed as environment
variables through \texttt{envFrom} or
\texttt{env.\allowbreak valueFrom.\allowbreak configMapKeyRef}, the values are
resolved at container startup and baked into the process environment. Subsequent
ConfigMap updates have no effect on running containers. This creates the silent
misconfiguration pattern: an operator updates a ConfigMap believing the change
is live, while running pods continue operating with the previous values
indefinitely.

When consumed as volume mounts, the kubelet performs an atomic symlink swap
on the \texttt{..data} directory within the projected volume on ConfigMap
update. The propagation delay is typically 10--90 seconds depending on the
kubelet's sync period configuration.

\section{Related Work}
\label{sec:related}

\subsection{Metrics-Based Observability}

Prometheus~\cite{b1} is the de facto standard for Kubernetes metrics collection,
providing powerful time-series query capabilities through PromQL with native
Kubernetes service discovery. However, Prometheus operates on numeric metrics
sampled at fixed intervals and does not have a native concept of discrete events
or causal relationships between them. Fundamentally, Prometheus cannot solve
the evidence horizon problem because it has no object snapshot model: it cannot
record the exact resource limits of a container at kill time, nor preserve
ConfigMap state at a specific past timestamp. Furthermore, Prometheus's
poll-based architecture creates the H5 blind spot described in
Section~\ref{sec:background}: pods whose entire lifetime falls within one
scrape interval generate zero time-series data. A Prometheus alert can notify
that a pod's restart count increased; it cannot reconstruct what caused the
failure, what configuration was in effect, or confirm the event occurred for
short-lived pods.

Thanos~\cite{b13} and Cortex~\cite{b14} extend Prometheus with long-term
storage and multi-cluster query federation, addressing metric retention but
not the structural absence of causal context or the sampling blind spot. The
evidence horizon problem persists regardless of storage duration: what is never
captured cannot be retained.

\subsection{Distributed Tracing}

Jaeger~\cite{b2} implements distributed request tracing, capturing causality
at the application request level as requests propagate through microservices.
These systems are fundamentally application-instrumented and do not observe
Kubernetes infrastructure events. The causal chain between a node memory
pressure condition and a container OOMKill is invisible to distributed tracing
systems because neither event originates from application instrumentation.
Distributed tracing cannot solve the evidence horizon problem because it
operates above the container runtime boundary.

OpenTelemetry~\cite{b4} provides a unified instrumentation standard for
metrics, logs, and traces, with growing support for Kubernetes infrastructure
signals through the OpenTelemetry Collector's Kubernetes receiver. However,
the OpenTelemetry Collector does not capture Kubernetes object state
transitions, construct causal edges between events, or address the evidence
horizon as a design requirement.

\subsection{Log Aggregation and Root Cause Analysis}

Elasticsearch~\cite{b3} and related log aggregation platforms aggregate
application and system logs into searchable indices. Log-based root cause
analysis~\cite{b6} and anomaly detection approaches have been extensively
explored. However, log aggregation cannot solve the evidence horizon problem
because it captures application output, not Kubernetes object state.

Loki~\cite{b15} extends the Grafana observability stack with log aggregation
designed for cloud-native environments. While Loki improves log accessibility
within the Kubernetes ecosystem, it shares the fundamental limitation of
log-based approaches: it captures what applications emit, not the Kubernetes
control plane state that the evidence horizon destroys.

\subsection{Kubernetes-Native Observability Tools}

The \texttt{kubectl} command-line tool~\cite{b5} provides direct access to the
Kubernetes API state but is inherently present-tense: it describes the current
state of objects, not historical states. Once a pod is deleted or its
\texttt{ContainerStatus} is overwritten, the previous state is inaccessible.
\texttt{kubectl debug}~\cite{b5} enables ephemeral container injection for
live diagnosis, but generates no persistent record of the debug session.

The Kubernetes audit log~\cite{b7} captures API server requests and can
reconstruct object state changes, but requires cluster-level configuration,
generates high log volume, and is not designed for causal query patterns.

\subsection{Backup, Recovery, and State Preservation}

Velero~\cite{b12} provides backup and restore capabilities for Kubernetes
workloads, capturing object state at scheduled intervals for disaster recovery.
While Velero preserves object state, it operates at scheduled backup intervals
rather than in response to specific event transitions, does not construct
causal edges between events, and is designed for recovery---not for real-time
diagnostic evidence preservation.

\subsection{Causal Inference in Distributed Systems}

Causal inference in distributed systems has been extensively
studied~\cite{b8,b9}. Vector clocks and happened-before
relationships~\cite{b9} provide formal foundations for establishing causal
ordering in distributed logs. OMA's causal edge construction is analogous to
happened-before relationships: if an \texttt{OOMKillEvidence} event $e_2$ is
observed for pod $P$ within 90 seconds of an \texttt{OOMKill} event $e_1$ for
the same pod, then $e_1 \rightarrow e_2$ in the happened-before sense. OMA
differs from Pearlian causal inference~\cite{b8} by encoding domain-specific
temporal constraints derived from Kubernetes semantics as first-class pattern
definitions, rather than inferring dependencies statistically.

Recent work on microservice failure diagnosis has applied causal graph models
to distributed system failures. CloudRCA~\cite{b10} demonstrates causal graph
construction for cloud platform root cause analysis. MicroScope~\cite{b16}
applies causal analysis to microservice performance degradation. These
approaches operate at the application and service mesh level; OMA operates at
the Kubernetes infrastructure level and addresses the evidence destruction
problem that precedes any higher-level causal analysis.

\subsection{AIOps and Container Orchestration Foundations}

AIOps platforms~\cite{b17} apply machine learning to IT operational data for
anomaly detection and incident correlation. The evidence horizon problem is
upstream of AIOps capabilities: if causal context is destroyed within 90
seconds of occurrence, no AIOps system can recover it retroactively. OMA
addresses the preservation layer that higher-level analysis depends on.

Burns \textit{et al.}~\cite{b11} describe the evolution from Borg and Omega
to Kubernetes, providing architectural context for the scheduling and lifecycle
management mechanisms that give rise to the evidence horizons formalized in
this work. The evidence horizon is an emergent consequence of Kubernetes design
principles~\cite{b11}: event-driven reconciliation, eventually consistent
object state, and garbage collection of ephemeral state.

\section{Evidence Horizon Taxonomy}
\label{sec:taxonomy}

\subsection{Formal Definition}

We define an \emph{evidence horizon} as a deterministic boundary, imposed by
the Kubernetes platform architecture, after which a specific category of
diagnostic context becomes permanently unrecoverable from the Kubernetes API
without prior preservation. Three properties characterize an evidence horizon:

\begin{enumerate}
\item \textbf{Determinism}: the destruction event occurs on a fixed schedule
or in response to a fixed platform trigger, independent of operator action.
\item \textbf{Irrecoverability}: once the horizon is crossed, the lost context
cannot be reconstructed from any remaining Kubernetes API state, regardless
of query latency or tooling sophistication.
\item \textbf{Specificity}: each horizon destroys a distinct category of
context; the destruction of one category does not imply the destruction of
others.
\end{enumerate}

This definition intentionally excludes data loss due to operator error,
cluster misconfiguration, or storage failure. Evidence horizons are structural
properties of the Kubernetes architecture that persist across all conformant
distributions and managed services.

\subsection{Taxonomy}

Table~\ref{tab:taxonomy} presents the five evidence horizons identified in
this work. Each horizon is characterized by its destruction trigger, the
category of data destroyed, and the OMA coverage provided by the corresponding
pattern encoder.

\begin{table}[!t]
\caption{Evidence Horizon Taxonomy}
\label{tab:taxonomy}
\centering
\setlength{\tabcolsep}{4pt}
\begin{tabular}{|p{0.04\linewidth}|p{0.14\linewidth}|p{0.13\linewidth}|p{0.15\linewidth}|p{0.22\linewidth}|p{0.13\linewidth}|p{0.07\linewidth}|}
\hline
\textbf{ID} & \textbf{Name} & \textbf{TTL / Trigger} & \textbf{Mechanism} &
\textbf{Data Destroyed} & \textbf{OMA Coverage} & \textbf{Pattern} \\
\hline
H1 & LastTermination-State Rotation &
$\sim$90\,s per restart &
Pod restart overwrites \texttt{LastTerminationState} &
Exit code, resource limits, ConfigMaps at kill time &
Full capture &
P001--P003 \\
\hline
H2 & Scheduler Event Pruning &
1\,hr / 1,000 events &
kube-apiserver TTL pruning &
Placement decisions, predicate failures &
Full capture &
P004 \\
\hline
H3 & Ephemeral Container State &
Immediate on exit &
API spec: no \texttt{lastState} field &
Exit code, duration, target context &
Full capture &
P005 \\
\hline
H4 & Kubelet Reconciliation Gap &
Node restart (15--60\,s) &
In-memory state not persisted &
Pending ops, probe state &
Theoretical &
--- \\
\hline
H5 & Scrape-Interval Blind Spot &
Per scrape interval &
Poll-based architecture &
Sub-interval pod lifetimes &
Structural immunity &
P001 \\
\hline
\end{tabular}
\end{table}

\subsection{H1: LastTerminationState Rotation}

\begin{figure}[t]
  \centering
  \includegraphics[width=\linewidth]{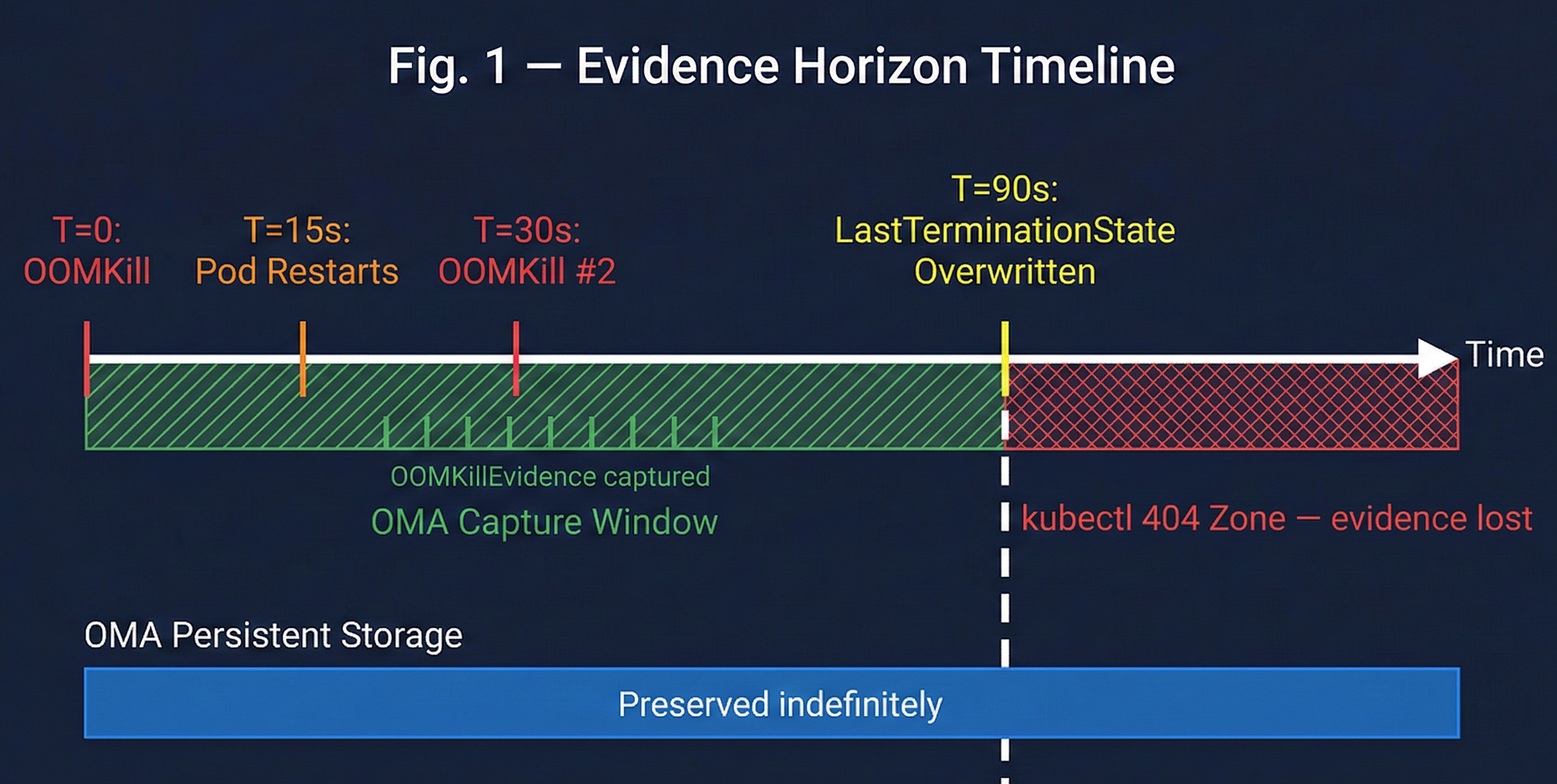}
  \caption{The H1 evidence horizon timeline. \texttt{LastTerminationState}
  is overwritten on each container restart, permanently discarding OOMKill
  forensic data. OMA captures \texttt{OOMKillEvidence} synchronously within
  the capture window; the operational memory store preserves the frozen state
  indefinitely after \texttt{kubectl} returns HTTP~404.}
  \label{fig1}
\end{figure}

H1 is the foundational evidence horizon introduced in~\cite{bv1}. The
\texttt{LastTerminationState.Terminated} field is overwritten on each container
restart, creating a $\sim$90-second window during which OOMKill forensic
evidence is accessible. Fig.~\ref{fig1} illustrates the H1 timeline: OMA
captures \texttt{OOMKillEvidence} synchronously within the capture window,
and the operational memory store preserves the frozen state indefinitely after
the pod is deleted and \texttt{kubectl} returns HTTP~404. Patterns P001, P002,
and P003 address H1.

\subsection{H2: Scheduler Event Pruning}

H2 introduces a second destruction mechanism operating on a longer but equally
deterministic schedule. The Kubernetes scheduler emits \texttt{FailedScheduling}
events containing structured predicate failure reasons and \texttt{Scheduled}
events recording the final placement decision. These events are pruned at the
one-hour TTL or at the 1,000-event cluster limit, whichever occurs first.

The diagnostic consequence of H2 is most severe in the cross-horizon case: a
pod is rejected from all nodes except one, placed on that node, and
subsequently OOMKills. After H2 pruning, the causal link between the
scheduler's constrained placement decision and the downstream OOMKill is
permanently severed. OMA's P004 pattern captures this link through a
cross-horizon causal edge (P004$\rightarrow$P001, conf=0.8), demonstrated empirically in Section~\ref{sec:evaluation}
(see Fig.~\ref{fig4}).

\subsection{H3: Ephemeral Container State}

H3 is structurally distinct from H1 and H2: the destruction is not timer-based
but specification-based. The \texttt{EphemeralContainerStatus} struct
explicitly excludes the \texttt{lastState} field that \texttt{ContainerStatus}
carries. This exclusion means that ephemeral container state is destroyed by
the platform at the moment of container exit, with no retention window.

OMA's P005 pattern captures the termination state at the \texttt{Terminated}
transition before the pod's next modification event overwrites the current
state. The explicit API boundary on log content capture is documented in the
pattern definition (Section~\ref{sec:architecture}).

Fig.~\ref{fig5} illustrates the evidence gap between what \texttt{kubectl}
returns and what OMA preserves after ephemeral container exit.

\subsection{H4: Kubelet Reconciliation Gap}

H4 identifies a fourth destruction mechanism arising from kubelet process
restart. Unlike H1--H3, which destroy persisted API state, H4 destroys
in-memory operational state that was never persisted to the Kubernetes API.
Full causal capture would require a kubelet-level integration beyond the
current OMA architecture. H4 is included in the taxonomy as a formally
identified horizon to guide future work.

\subsection{H5: Scrape-Interval Blind Spot}

H5 is a structural property of poll-based observability rather than a
Kubernetes API feature. OMA's watch-based architecture provides structural
immunity: every pod lifecycle event is delivered at the moment of occurrence,
independent of any scrape schedule. No new pattern encoder is required. H5 is
an emergent advantage of the OMA architecture, validated empirically in
Section~\ref{sec:evaluation}.

\subsection{Taxonomy Summary}

The five horizons form two structural groups. H1, H2, and H3 destroy
Kubernetes API state that previously existed. H4 and H5 prevent observability
data from reaching any persistent store. OMA addresses the first group with
pattern-specific watchers and the operational memory store, and exploits its
event-driven architecture to provide structural immunity to the second group.

\section{Operational Memory Architecture}
\label{sec:architecture}

\subsection{Design Principles}

OMA is designed around three principles carried forward from~\cite{bv1} and
extended to the full taxonomy. First, \emph{evidence preservation before
rotation}: the system must capture relevant context within each evidence
horizon, not after. Second, \emph{operational causality construction}: OMA
constructs causal edges using domain-specific temporal constraints and
Kubernetes semantics, analogous to happened-before relationships~\cite{b9},
rather than statistical causal inference in the Pearlian sense~\cite{b8}.
Third, \emph{query-first design}: the system is optimized for specific query
patterns that address real operational questions.

\subsection{Extended Four-Layer Architecture}

\begin{figure}[t]
  \centering
  \includegraphics[width=\linewidth]{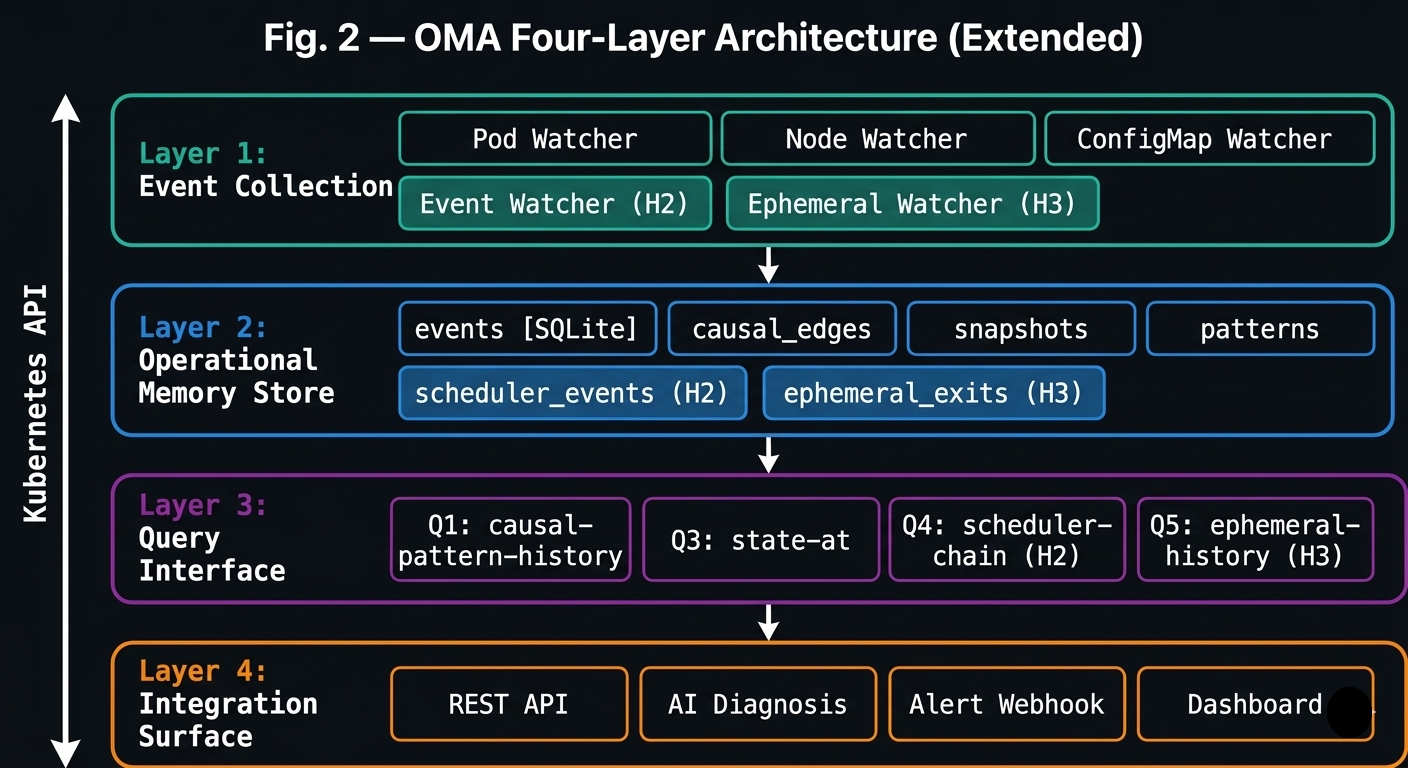}
  \caption{Extended OMA four-layer architecture. Layer~1 adds
  \texttt{EventWatcher} (H2) and \texttt{EphemeralWatcher} (H3) to the
  original three watchers. Layer~2 adds \texttt{scheduler\_\allowbreak events}
  and \texttt{ephemeral\_\allowbreak exits} tables. Layer~3 adds Q4 and Q5
  canonical queries for cross-horizon provenance and ephemeral exit history.}
  \label{fig2}
\end{figure}

The OMA four-layer architecture is extended to address H2 and H3, as
illustrated in Fig.~\ref{fig2}.

\textit{Layer~1---Event Collection.} The Go collector now comprises five
concurrent watchers. \texttt{NodeWatcher}, \texttt{PodWatcher}, and
\texttt{ConfigMapWatcher} are carried from~\cite{bv1} and address H1.
\texttt{EventWatcher} (new) captures \texttt{FailedScheduling},
\texttt{Scheduled}, and \texttt{Preempting} events from
\texttt{default-scheduler} before kube-apiserver TTL pruning (H2).
\texttt{EphemeralWatcher} (new) inspects \texttt{Ephemeral\allowbreak Container\allowbreak Statuses}
on each pod \texttt{Modified} event, firing on the
\texttt{Running$\rightarrow$Terminated} transition (H3).

\textit{Layer~2---Operational Memory Store.} The SQLite schema is extended
with two new tables. The \texttt{scheduler\_events} table stores P004 records
with parsed predicate failures (JSON array), pruning risk classification, and
TTL expiry timestamp. The \texttt{ephemeral\_exits} table stores P005 records
with exit code, exit class, duration, target container, and
\texttt{log\_api\_boundary}. Both tables use the same event \texttt{id} as
primary key, enabling JOIN queries against the main \texttt{events} table.

\textit{Layer~3---Query Interface.} Q1--Q3 from~\cite{bv1} are extended with
Q4 (\texttt{scheduler-chain}) for P004 cross-horizon provenance and Q5
(\texttt{ephemeral-history}) for P005 session history.

\textit{Layer~4---Integration Surface.} Unchanged from~\cite{bv1}.

\subsection{Causal Pattern Encoding}

\textit{Pattern P004 (Scheduler Decision Provenance)} encodes:
\texttt{FailedScheduling} [precursor, 3600\,s window] $\rightarrow$
\texttt{Scheduled} [trigger] $\rightarrow$ \texttt{OOMKill} [effect,
cross-horizon, conf=0.8]. The 3600-second window encodes the H2 TTL directly.
The cross-horizon edge carries confidence 0.8 because the scheduler's placement
decision is a contributing causal factor but not the sole determinant of an
OOMKill.

\textit{Pattern P005 (Ephemeral Container Evidence Loss)} encodes:
\texttt{EphemeralContainerTerminated} [trigger, immediate]. The single-step
pattern reflects H3's instantaneous nature. Each P005 record is a standalone
capture; no prior-session cross-cycle edge is possible because
\texttt{EphemeralContainerStatus} has no \texttt{lastState} field.

\subsection{Cross-Horizon Causal Edge Construction}

The cross-horizon P004$\rightarrow$P001 edge is constructed during ingest when
a \texttt{Scheduled} event is processed: the ingest layer queries the
\texttt{events} table for an \texttt{OOMKill} record for the same pod and
namespace, and inserts an edge with \texttt{pattern\_id=P004},
\texttt{confidence=0.8}, and
\texttt{edge\_type=cross\_pattern\_P004\_P001}. The existing
\texttt{cause\_event\_id} and \texttt{effect\_event\_id} foreign keys in
\texttt{causal\_edges} accommodate this edge without schema modification.

\subsection{Fault Tolerance and API Boundaries}

The at-least-once delivery model and \texttt{INSERT OR IGNORE} idempotency
from~\cite{bv1} apply to all five patterns without modification. Two API
boundaries are explicitly documented in P005: log content is not capturable
after container exit (\texttt{log\_content=NOT\_CAPTURABLE\_VIA\_API}), and
\texttt{source.component} is not a supported watch field selector for
\texttt{Event} objects (client-side filtering applied).

\section{Implementation}
\label{sec:implementation}

\subsection{Go Collector Extensions}

The collector is implemented in Go~1.21 using \texttt{k8s.io/client-go}.
Two new watchers follow the same goroutine-per-watcher pattern with
context-based cancellation and automatic watch reconnection on channel closure.

\texttt{EventWatcher} subscribes to
\texttt{CoreV1().Events(namespace).Watch()} with empty \texttt{ListOptions}.
Filtering to scheduler events is applied client-side by checking
\texttt{Event.Source.Component == "default-scheduler"}. On each qualifying
event, the watcher emits a \texttt{CausalEvent} with
\texttt{EventType="SchedulerEvent"}, \texttt{PatternID="P004"}, and a payload
containing the full predicate failure message, event age, pruning risk
classification, and computed \texttt{evidence\_expires} timestamp.

\texttt{EphemeralWatcher} subscribes to
\texttt{CoreV1().Pods(namespace).Watch()}. On each \texttt{Modified} event,
it inspects \texttt{Pod.Status.EphemeralContainerStatuses}. A per-container
key (\texttt{namespace/pod/container-name}) prevents duplicate emissions on
repeated \texttt{Modified} events for the same exit. The exit class is derived
from exit code and reason: \texttt{CLEAN} (0), \texttt{OOM}
(\texttt{OOMKilled}), \texttt{SIGKILL} (137), \texttt{SIGTERM} (143), or
\texttt{ERROR} (any other non-zero code). The \texttt{main.go} entry point
adds two goroutines and increases the error channel buffer from 3 to 5.

\subsection{Pattern Registration}

Two new pattern definitions are added to the \texttt{patterns} package
following the existing \texttt{CausalPattern} struct. Each registers itself
in the global \texttt{AllPatterns} map via \texttt{init()}.

\subsection{Storage Layer Extensions}

\texttt{schema\_v2.sql} creates \texttt{scheduler\_events} and inserts the
P004 pattern row. \texttt{schema\_v3.sql} creates \texttt{ephemeral\_exits}
and inserts P005. Both use \texttt{IF NOT EXISTS} throughout and are
idempotent. The \texttt{ingest\_v2.py} module integrates via two one-line
additions to \texttt{ingest.py}: one call at the end of
\texttt{\_insert\_event()} and one at the end of \texttt{\_build\_edges()}.
No existing logic is modified.

\subsection{Timestamp Handling}

Kubernetes API timestamps include UTC offset suffixes
(e.g., \texttt{2026-04-17T11:29:01.789683-05:00}). The ingest layer
normalizes these to bare ISO~8601 strings before SQLite comparisons,
consistent with the approach in~\cite{bv1} for P001 edge construction.

\section{Evaluation}
\label{sec:evaluation}

\subsection{Experimental Setup}

We evaluate the extended OMA through six scenario validations spanning H1--H3
and H5, a 30-run statistical latency analysis carried forward from~\cite{bv1},
and a concurrent stress evaluation. Environment~1 is a local Minikube cluster
(Kubernetes~1.31, 3~nodes) on Apple M-series hardware. Environment~2 is an
Azure Kubernetes Service cluster (version~1.32.10, 2~Standard\_B2s nodes) in
\texttt{eastus}. All raw JSONL files, SQLite databases, and scenario trigger
scripts are committed under \texttt{docs/poc-results/} for independent
verification.

\subsection{H1 Scenario Validation (P001, P002, P003)}

H1 validation results from~\cite{bv1} are carried forward. Table~\ref{tab:h1}
summarizes the scenario validation across both environments.

\begin{table}[!t]
\caption{H1 Scenario Validation Results}
\label{tab:h1}
\centering
\setlength{\tabcolsep}{5pt}
\begin{tabular}{|p{0.30\linewidth}|p{0.17\linewidth}|p{0.17\linewidth}|p{0.20\linewidth}|}
\hline
\textbf{Metric} & \textbf{Minikube R1} & \textbf{Minikube R2} & \textbf{AKS 1.32.10} \\
\hline
Total events (P001)   & 30    & 31    & 20   \\
OOMKill events        & 6     & 6     & 4    \\
OOMKillEvidence       & 16    & 16    & 10   \\
Causal edges          & 13    & 13    & 8    \\
Exit code             & 137   & 137   & 137  \\
Edge confidence       & 1.0   & 1.0   & 1.0  \\
P002 captured         & Yes   & ---   & Yes  \\
P003 captured         & Yes   & ---   & Yes  \\
\hline
\end{tabular}
\end{table}

\begin{figure}[t]
  \centering
  \includegraphics[width=\linewidth]{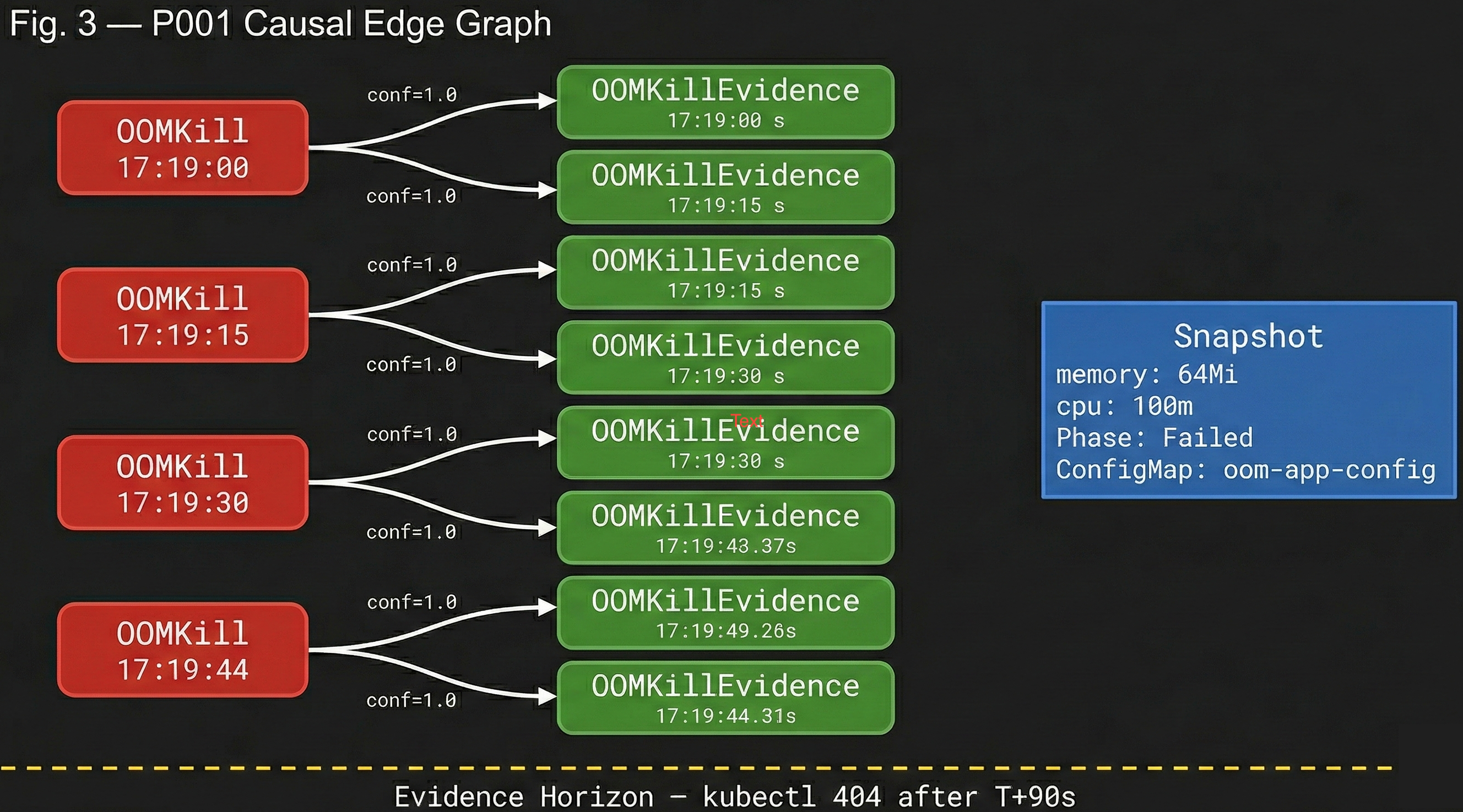}
  \caption{P001 causal edge graph from the AKS run
  (\texttt{aks-nodepool1-78296979-vmss000000}). Four OOMKill events (red)
  are linked to eight \texttt{OOMKill\allowbreak Evidence} events (green)
  via directed causal edges with confidence~1.0. A point-in-time snapshot
  (blue) preserves the complete pod state after deletion. The dashed line
  marks the H1 evidence horizon beyond which \texttt{kubectl} returns
  HTTP~404.}
  \label{fig3}
\end{figure}

\subsection{H2 Scenario Validation (P004)}

We deployed an unschedulable pod (\texttt{memory: 999Gi}) triggering
\texttt{FailedScheduling} events, and a schedulable pod with a 64\,Mi limit
producing a \texttt{Scheduled} event followed by OOMKill cycles. Minikube was
started with \texttt{--extra-config=apiserver.event-ttl=2m} to make the H2
pruning observable within the scenario runtime. Table~\ref{tab:h2} reports the
results.

\begin{table}[!t]
\caption{H2 Scenario Validation Results (P004)}
\label{tab:h2}
\centering
\setlength{\tabcolsep}{5pt}
\begin{tabular}{|p{0.52\linewidth}|p{0.32\linewidth}|}
\hline
\textbf{Metric} & \textbf{Value} \\
\hline
SchedulerEvents captured       & 2 (FailedScheduling, Scheduled) \\
FailedScheduling message       & 0/3 nodes: 3 Insufficient memory \\
Scheduled placement            & opscart-m02 \\
P004 sequence edges            & 1 (conf=1.0) \\
P004$\rightarrow$P001 edge     & 1 (conf=0.8, cross-horizon) \\
After 2\,min TTL: kubectl      & No resources found \\
After 2\,min TTL: OMA          & Full chain preserved \\
\hline
\end{tabular}
\end{table}

The \texttt{FailedScheduling} message was parsed into structured predicate
failures and stored as a JSON array. After TTL expiry, \texttt{kubectl get
events -n oma-scheduler} returned \texttt{No resources found}; the OMA
\texttt{scheduler\_events} table retained the complete record. The
cross-horizon causal edge (P004$\rightarrow$P001, conf=0.8) was confirmed in
\texttt{causal\_edges} with
\texttt{edge\_type=cross\_pattern\_P004\_P001}.

\begin{figure}[t]
  \centering
  \includegraphics[width=\linewidth]{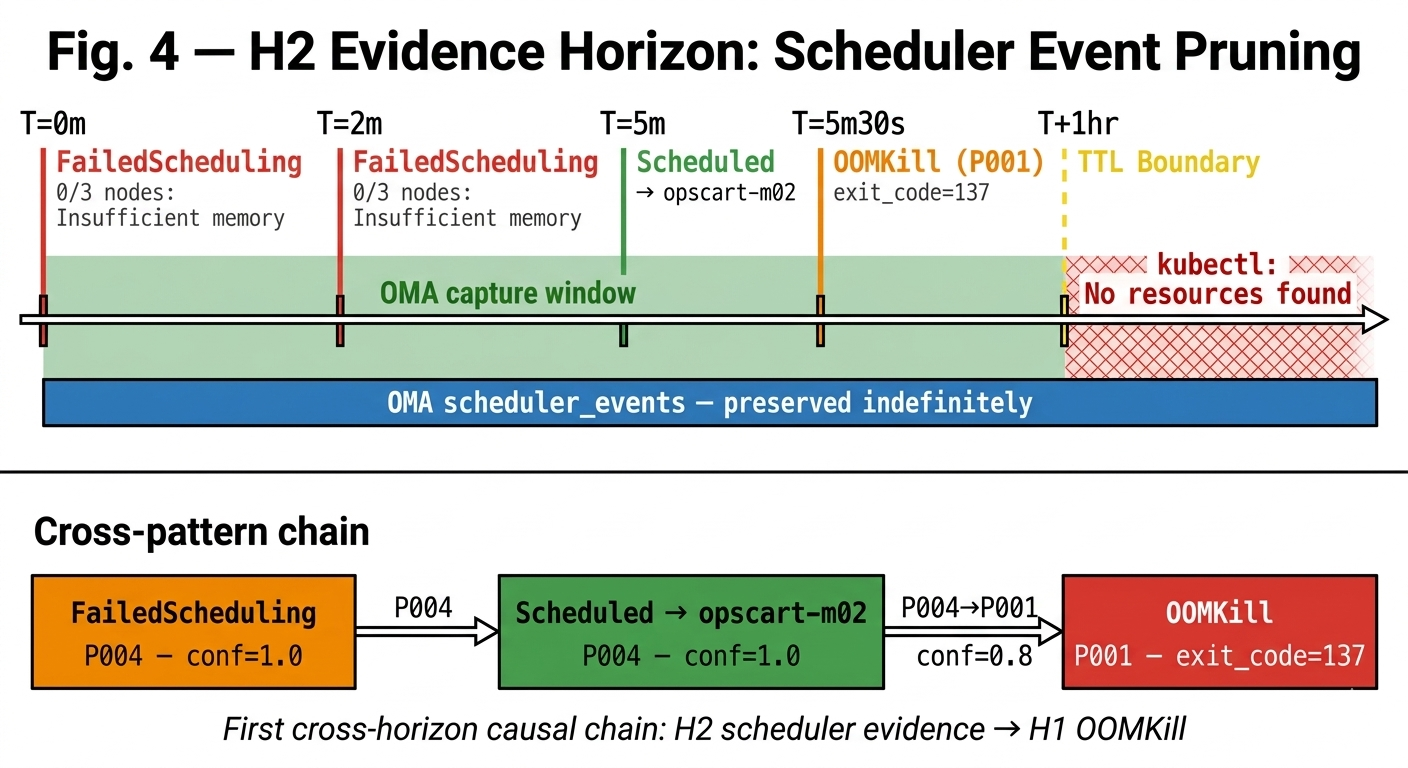}
  \caption{H2 evidence horizon: scheduler event pruning timeline and
  cross-horizon causal chain. \texttt{Failed\allowbreak Scheduling} and
  \texttt{Scheduled} events are captured by OMA before the kube-apiserver
  TTL boundary (dashed yellow). After TTL expiry, \texttt{kubectl} returns
  no resources; OMA preserves the complete chain including the cross-horizon
  P004$\rightarrow$P001 edge (conf=0.8) linking the placement decision to
  the downstream OOMKill.}
  \label{fig4}
\end{figure}

\subsection{H3 Scenario Validation (P005)}

We attached ephemeral debug container \texttt{oma-debug-1776446626} to pod
\texttt{ephemeral-target} (node \texttt{opscart-m02}) via \texttt{kubectl
debug --target=app}. The container ran for 10 seconds and exited with
code~42. Table~\ref{tab:h3} reports the capture results.

\begin{table}[!t]
\caption{H3 Scenario Validation Results (P005)}
\label{tab:h3}
\centering
\setlength{\tabcolsep}{5pt}
\begin{tabular}{|p{0.45\linewidth}|p{0.40\linewidth}|}
\hline
\textbf{Field} & \textbf{Value} \\
\hline
\texttt{container\_name}    & oma-debug-1776446626 \\
\texttt{target\_container}  & app \\
\texttt{exit\_code}         & 42 \\
\texttt{exit\_class}        & ERROR \\
\texttt{duration\_seconds}  & 10.0 \\
\texttt{node\_name}         & opscart-m02 \\
\texttt{log\_content}       & NOT\_CAPTURABLE\_VIA\_API \\
kubectl \texttt{lastState}  & (empty --- no field in spec) \\
kubectl logs (after exit)   & Error: container not found \\
\hline
\end{tabular}
\end{table}

\begin{figure}[t]
  \centering
  \includegraphics[width=\linewidth]{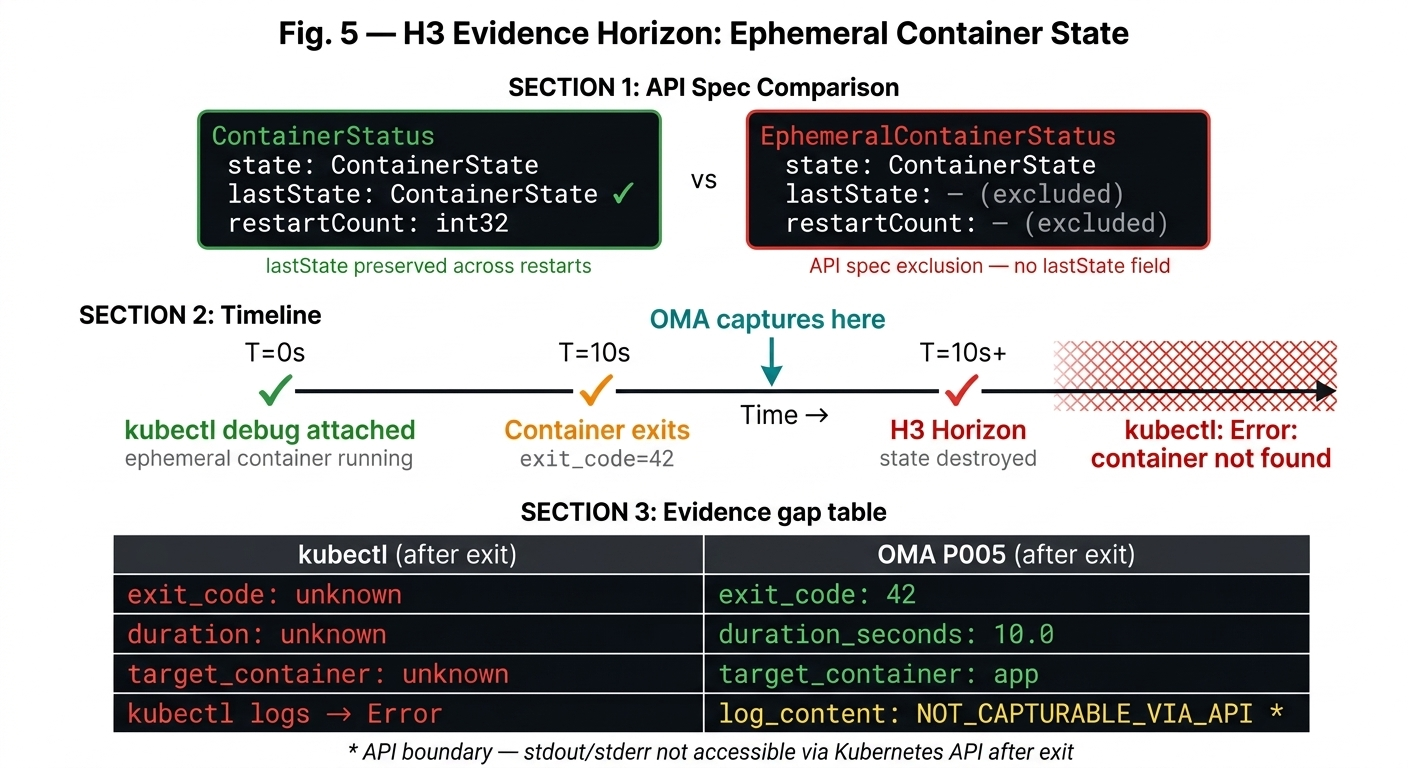}
  \caption{H3 evidence horizon: ephemeral container state loss.
  \texttt{Ephemeral\allowbreak ContainerStatus} explicitly excludes the
  \texttt{lastState} field present in \texttt{Container\allowbreak Status}
  (top). After container exit, \texttt{kubectl} returns no termination
  record; OMA P005 captures exit code, duration, and target container
  context at the \texttt{Terminated} transition (middle). The evidence gap
  table (bottom) shows fields preserved by OMA that are unrecoverable from
  the Kubernetes API.}
  \label{fig5}
\end{figure}

The H3 evidence gap is confirmed: \texttt{kubectl describe pod} showed no
\texttt{lastState} entry for the ephemeral container, consistent with the
API specification exclusion. The OMA \texttt{ephemeral\_exits} table retained
the complete P005 record with all fields populated.

\subsection{H5 Scenario Validation}

Ghost pod \texttt{ghost-pod} (namespace \texttt{oma-sampling}) OOMKilled at
T+5\,s with an observed lifetime of 6\,s on node \texttt{opscart-m03}. With
a default Prometheus scrape interval of 15\,s, the pod's lifetime falls
entirely within one scrape gap. The OMA \texttt{PodWatcher} captured 2
\texttt{OOMKill} P001 events at occurrence with \texttt{exit\_code=137} and
\texttt{memory\_limit=64Mi}. The Prometheus result is confirmed by the
architectural argument: with pod lifetime 6\,s $<$ scrape interval 15\,s,
the PromQL query
\texttt{container\_cpu\_usage\_seconds\_total\{pod="ghost-pod"\}} returns an
empty result set---a deterministic consequence of the sampling architecture,
not dependent on Prometheus configuration.

\subsection{Statistical Latency Analysis (30 Runs)}

The 30-run P001 latency analysis from~\cite{bv1} is carried forward.
Table~\ref{tab:latency} reports the bimodal distribution across 242 edges.

\begin{table}[!t]
\caption{Causal Edge Construction Latency (30 Runs, 242 Edges)}
\label{tab:latency}
\centering
\setlength{\tabcolsep}{5pt}
\begin{tabular}{|p{0.33\linewidth}|p{0.09\linewidth}|p{0.12\linewidth}|p{0.12\linewidth}|p{0.12\linewidth}|}
\hline
\textbf{Edge Class} & \textbf{Count} & \textbf{Min} & \textbf{Mean} & \textbf{Max} \\
\hline
Intra-cycle ($<$100\,ms) & 88  & 0.089\,ms & 0.702\,ms  & 2.607\,ms  \\
Cross-cycle ($\geq$100\,ms) & 154 & 903\,ms & 12,708\,ms & 31,454\,ms \\
\hline
\end{tabular}
\end{table}

The intra-cycle mean of 0.702\,ms ($\sigma=0.31$\,ms) confirms synchronous
evidence capture well within the H1 boundary. Cross-cycle latency reflects
CrashLoopBackOff restart interval timing, not processing delay.

\subsection{Stress Evaluation}

Table~\ref{tab:stress} confirms linear event scaling and flat memory
consumption under concurrent load, carried forward from~\cite{bv1}.

\begin{table}[!t]
\caption{Stress Evaluation: Concurrent OOMKill Pods}
\label{tab:stress}
\centering
\setlength{\tabcolsep}{5pt}
\begin{tabular}{|p{0.10\linewidth}|p{0.14\linewidth}|p{0.17\linewidth}|p{0.12\linewidth}|p{0.15\linewidth}|}
\hline
\textbf{Pods} & \textbf{Events} & \textbf{Events/sec} & \textbf{Edges} & \textbf{RAM (MB)} \\
\hline
5  & 95  & 0.77 & 51  & 7.9 \\
10 & 175 & 1.43 & 90  & 8.2 \\
20 & 355 & 2.86 & 197 & 8.8 \\
\hline
\end{tabular}
\end{table}

\subsection{Extended Capability Comparison}

Table~\ref{tab:comparison} extends the capability comparison from~\cite{bv1}
to include H2, H3, and H5 capabilities.

\begin{table}[!t]
\caption{Extended Capability Comparison}
\label{tab:comparison}
\centering
\setlength{\tabcolsep}{4pt}
\begin{tabular}{|p{0.38\linewidth}|p{0.13\linewidth}|p{0.16\linewidth}|p{0.13\linewidth}|}
\hline
\textbf{Capability} & \textbf{kubectl} & \textbf{Prometheus} & \textbf{OMA} \\
\hline
OOMKill evidence after restart  & $<$90\,s  & No        & Indefinite \\
Resource limits at kill time    & $<$90\,s  & Approx.   & Exact, frozen \\
ConfigMap in effect at failure  & No        & No        & Refs + hash \\
Stale env var detection         & No        & No        & Yes (P002) \\
ConfigMap propagation latency   & No        & No        & Yes (P003) \\
State of deleted objects        & 404       & Partial   & Yes (Q3) \\
Causal edges between events     & No        & No        & Yes \\
Pattern recurrence detection    & No        & Via rules & Yes (Q2) \\
Scheduler placement rationale   & $<$1\,hr  & No        & Yes (P004) \\
Ephemeral container exit code   & No field  & No        & Yes (P005) \\
Sub-interval pod events         & 404       & 0 points  & Yes (P001) \\
\hline
\end{tabular}
\end{table}

\section{Discussion}
\label{sec:discussion}

\subsection{Limitations}

\textit{H4 implementation boundary.} Full causal capture of the kubelet
reconciliation gap requires a kubelet-level integration outside the current
OMA architecture. OMA's \texttt{NodeWatcher} detects the
\texttt{NodeNotReady$\rightarrow$NodeReady} transition and records the gap
duration, but cannot recover in-memory operational state lost during the
kubelet restart. This is the primary future work direction.

\textit{H5 Prometheus validation.} The H5 sampling blind spot is validated
through the architectural argument (pod lifetime 6\,s $<$ scrape interval
15\,s) and OMA's empirical capture of the OOMKill event. A full empirical
comparison with a Prometheus installation was not performed; the architectural
argument is deterministic and environment-independent.

\textit{NodeMemoryPressure edge.} The P004$\rightarrow$P001 cross-horizon edge
carries confidence 0.8. The \texttt{NodeMemoryPressure$\rightarrow$OOMKill}
edge (conf=0.9, defined in P001) was not observed because the test node's
4\,GB RAM is not exhausted by a single 64\,Mi pod OOMKill. Both edge types
are implemented; the NodeMemoryPressure scenario requires node-level memory
exhaustion across multiple pods.

\textit{Namespace scope.} The collector is namespace-scoped. Cluster-wide
deployment requires a \texttt{ClusterRole} with watch permissions or
per-namespace collector instances.

\textit{Statistical evaluation scope.} The 30-run latency analysis was
conducted on Minikube on Apple M-series hardware. A full statistical
evaluation on AKS is consistent with single-run AKS results but is deferred
to future work.

\subsection{RBAC and Security Considerations}

\texttt{EventWatcher} requires \texttt{get}, \texttt{list}, and \texttt{watch}
verbs on \texttt{Event} objects. \texttt{EphemeralWatcher} requires the same
verbs on \texttt{Pod} objects, which the existing \texttt{PodWatcher} already
holds. No additional cluster-level permissions are required for
namespace-scoped deployment.

\subsection{Production Deployment Considerations}

The two new watchers add no measurable memory overhead beyond the per-watcher
goroutine stack ($\sim$8\,KB each) and the \texttt{lastSeen} map in
\texttt{EphemeralWatcher}, which grows proportionally to observed ephemeral
containers rather than event volume. The streaming JSONL model confirmed at
8.8\,MB RAM under 20 concurrent OOMKill pods applies equally to the extended
collector. OMA's immutable event log constitutes an auditable record of
scheduler decisions, container failure reasons, and debug session activity
suitable for compliance environments.

\section{Conclusion}
\label{sec:conclusion}

We have presented an extended Operational Memory Architecture that formalizes
Kubernetes evidence destruction as a taxonomy of five structurally distinct
horizons and extends OMA's causal preservation capabilities to address three
of them empirically and two theoretically. The evidence horizon taxonomy
(H1--H5) provides a systematic framework for reasoning about diagnostic
context loss in Kubernetes clusters, distinguishing between API state
destruction (H1--H3), in-memory state loss (H4), and architectural sampling
blind spots (H5).

The two new causal patterns demonstrate that evidence horizons operate
independently and compound in production failures. P004 captures scheduler
placement rationale before the H2 pruning boundary and constructs the first
cross-horizon causal chain linking scheduling decisions to downstream OOMKill
failures. P005 addresses the H3 API specification exclusion that makes
ephemeral container debug sessions leave no persistent forensic record.
Together with the H5 structural immunity demonstrated through the 6-second
ghost pod experiment, the extended OMA confirms that an event-driven
preservation architecture provides systematic advantages over poll-based
and API-query-based approaches across all five evidence horizons.

The original performance characteristics are preserved: intra-cycle causal
edge construction at mean 0.702\,ms, linear event scaling to 2.86 events/sec
under 20 concurrent OOMKill pods, and flat 8.8\,MB RAM consumption. The
extended architecture introduces no measurable overhead relative to the
original three-watcher design.

The complete implementation, all scenario trigger scripts, raw JSONL event
logs, SQLite databases, and query outputs are released as open-source software
at \url{https://github.com/opscart/k8s-causal-memory} for independent
verification and replication.

\appendices

\section*{Acknowledgments}

The author thanks the open-source Kubernetes and Go communities whose
tooling made this work possible.


\end{document}